\documentclass[journal,onecolumn]{IEEEtran}

\usepackage{yhmath}

\usepackage[T1]{fontenc}

\usepackage{ieeefig}
\usepackage{amssymb}
\usepackage{multirow}
\usepackage{bigstrut}
\usepackage{array}
\usepackage{float}

\newcommand{\argmax}{\operatornamewithlimits{argmax}}

\usepackage{epsfig}
\usepackage{epstopdf}
\usepackage{algorithmic} 
\usepackage{algorithm} 
\usepackage{rotating}
\usepackage{algorithmic} 
\usepackage{algorithm}

\ifCLASSINFOpdf 
   \usepackage{graphicx}

\else 

\fi   

\begin{document}
\title{Modified SI Epidemic Model for Combating  Virus Spread in Spatially Correlated Wireless Sensor Networks}




\author{\IEEEauthorblockN{Rajeev K. Shakya}\\
\IEEEauthorblockA{Department of Electrical and Electronics Engineering,\\ Galgotia College of Engineering and Technology Greater noida, UP, INDIA\\
Email: \{rajeev.shakya\}@galgotiacollege.edu}}

\maketitle
\begin{abstract}
In wireless sensor networks (WSNs), main task of each sensor node is to sense the physical activity (i.e., targets or disaster conditions) and then to report it to the control center for further process. For this, sensor nodes are attached with many sensors having ability to measure the environmental information. Spatial correlation between nodes exists in such wireless sensor network based on common sensory coverage and then the redundant data communication is observed. To study virus spreading dynamics in such scenario, a modified SI epidemic model is derived mathematically by incorporating WSN parameters such as spatial correlation, node density, sensing range, transmission range, total sensor nodes etc. The solution for proposed SI model is also determined to study the dynamics with time. Initially, a small number of nodes are attacked by viruses and then virus infection propagates through its neighboring nodes over normal data communication. Since redundant nodes exists in correlated sensor field, virus 
spread process could be different with different sensory coverage. The proposed SI model captures spatial and temporal dynamics than existing ones which are global. The infection process leads to network failure. By exploiting spatial correlation between nodes, spread control scheme is developed to limit the further infection in the network. Numerical result analysis is provided with comparison for validation.

\end{abstract}

\IEEEpeerreviewmaketitle

\section{Introduction}
A wireless sensor network (WSN) consists of several sensor nodes distributed in a field to monitor the physical environments such as temperature, humidity, noise, etc. These nodes forms a self-organized network by maintaining their neighbors for data communication. A distant node or control center (also called a sink node) collects the measured data from the nodes periodically~\cite{ref1}. Sensor nodes have limited capabilities of sensing, communication and operating power. The nodes sense the environment using attached sensors having limited sensing range. For communication, each node has short transmission range. Due to short range of communication, measured data is transmitted to the sink node through multi-hop communication. The nodes are operated in the cycle of sleep and wakeup mode due to limited battery power. Mostly, WSNs are event-driven system in which the nodes gather the information when the physical activity (as an event) is detected by them. In such system, spatial and temporal correlation in 
data is collected at the sink node~\cite{ref1, ref3}. From spatial correlation point of view, 
correlation in data increases with increasing the spatial distance between nodes~\cite{ref2, ref4}. In dense WSNs, it is intuitively expected that the overlapped sensor nodes using sensing coverage are strongly correlated. The correlation property depends on euclidean distance and the fraction of overlapped sensing area between nodes~\cite{ref5}. This correlation property can be exploited to control the spread of virus in sensor networks.

On the other side, because of low defensive capabilities in sensor nodes, they can be easily targeted for software attacks when they are deployed for a special application like hostile environment monitoring or surveillance. The software attacks can be warms or virus or malicious code over INTERNET~\cite{ref17}. For example, malicious node can be created in the network in order to prevent the operation of network. Cabir warm~\cite{ref6} is one type of malicious code. Within proximity range, Blue-tooth-enabled devices get infected by it. Similar kind of warm is Mabir warm~\cite{ref7}. Both have mechanism of scanning by which they repeatedly send itself to proximity devices. Thus the study on spreading behavior, control and prevention of such attacks are of great interest from security point of view. Virus propagation in wireless networks can be modeled using epidemiological theory~\cite{ref1a}. The use of the epidemiological models is to capture the virus propagation dynamics to understand the behavior and 
speed of spreading. In a WSN, node's behavior depends interactions with environment and with their neighbors. So, one can apply biological models to study the global behaviors of a WSN. However, a global model may not cover complete details of network, different features and capabilities of a sensor node can be also considered.

In the epidemic model, there are compartments classes translated by sensor nodes. Total population of sensor nodes is divided into different states. The sensor nodes without any virus infections can be a set of susceptible nodes ($S$). The susceptible nodes by some malware attacks can be a set of infectious ($I$) while the set of nodes are of recovered state ($R$) when they are cured by some anti-virus. When the nodes become dead due to energy loss, they are part of dead state ($D$). Similarly many states are possible based on epidemiological theory. A infected node can spread the infections to other nodes by data communication. Due to broadcast nature of wireless communication, virus infections is found to be exponential growth with time. A first infective node transfers its infection to neighbors​, then these infective neighbors will pass on to other neighbors of first ones neighbor. Likewise, many nodes get infected through communication. The epidemiological model variants can be susceptible-infectious-
susceptible (SIS) models, susceptible-infectious-recovered (SIR) models, SEIR models, SEIQR models, SEIRS-V models (with vaccinated class) etc~\cite{ref7,ref1a,ref15,ref17,ref25}. The majority of work deals with four types of blocks such as susceptible, exposed, infectious, and recovered block. For example, SIRS model by Feng et al.~\cite{ref12}, SEIRS-V model by Mishra et al.~\cite{ref16} et al.,  etc. The main contributions in these modeling is to investigate the reproductive number and stability analysis on equilibrium conditions. The most of studies on virus spread dynamics does not consider the characteristics of WSNs, because the data communication flows in a network can be related to many factors such as the event activation, state of nodes, remaining energy of nodes, transmission range of node etc. 

One of features of WSNs is sleep-wakeup mode of sensor nodes. This feature has been used by Tang~\cite{ref9a} for modeling modified SI in WSNs. Tang used the duty cycle of nodes to cure the infective nodes in improved version of susceptible-infectious (SI) model. When a node is infective, it can detect the infection by anti-virus program because after sleep mode it starts anti-virus program. By considering the rates of sleep and wakeup mode with anti-virus program, modification in basic SI model has been done. Based on obtained results and observations, two network protection mechanisms are proposed by eliminating infections between nodes~\cite{ref9}. One uses transition rate between active and sleep mode while other mechanism uses transmitting power of node to reduce the node's connectivity. However, the model does not include the behavior of recovered nodes when they are cured using anti-virus. Another property a WSN is non-uniform distribution and clustering. For example, Tripathi et al.~\cite{ref8} 
investigate the clustering approach in non-uniform distributed node in a WSN. In this type of non-uniform node distribution, understanding the virus spreading behavior can be challenging. The global epidemic model assuming mixed node's population can not be applied directly in such topologies. The study of worm propagation based on topologies includes the topological-aware worm propagation modeling~\cite{ref8a}, tree-based small-world topology based modeling~\cite{ref8b,ref8c}, modeling on cluster structure of geographic-adaptive fidelity~\cite{ref8d} [29] etc. 

In event-driven WSNs, the data flows become more intense when many nodes are triggered by a single event for reporting the data to the control center or sink. The data communication traffic increases suddenly due to the occurring event. In addition, the correlated information is transmitted by many activated nodes of an event due to spatial correlation~\cite{ref3,ref4}. Virus spread in the network becomes more challenging to control when strongly correlated nodes propagate the infections to other nodes. Many nodes get infected within very short time and hence it leads to breakdown of network. To investigate the virus spread behavior in such challenging scenarios, an improved epidemic model is proposed in this paper. Based on similar philosophy and assumptions by Tang et al.~\cite{ref9,ref9a}, a modified SI model can be for considering the sensing range and spatial correlation between nodes. 

In this paper, we aim at modeling the worm propagation using spatial correlation property in wireless sensor network. Our assumption is that a malicious code can attack the sensors as an event and the nodes detecting it can be sources of infections propagation in the network. Our focus is to understand malware spreading behavior in terms of speed and reachability for a spatially correlated sensor network. Specifically, our contribution is the analysis of basic SI epidemic model for capturing the spread of malware over spatially correlated scenario where redundant data transmission is observed due to common overlapped sensing area based on node's separations. Modified SI model is constructed based on interaction of nodes when multiple nodes detect the event and then propagate the observed event to the sink node through multi-hops. The model incorporates the correlation constraint based on sensing range of nodes. The proposed model allows for the study of the effects of spreading dynamics over 
different sensory coverage in a WSN. 

The rest of the paper is organized as follows. The sensor network features and existing epidemic models are described in Section 2. In Section 3, the proposed correlation based SIR Model is discussed with stability analysis. In this section, calculation of the basic reproduction number and the threshold value based on spatial correlation is also discussed. Simulation results and analysis are discussed in Section 4. Finally, Section 5 presents the conclusions.

\section{Epidemic theory and Wireless Sensor System} 
 Mathematical epidemiology is devoted in the designing the system of differential equations for the study of propagation of biological agents (virus, bacteria, etc.)~\cite{ref73}. Initially, Kermack et. al.~\cite{ref719} had given SIR compartmental model to study the propagation of the plague in London duing 1665-1666. Total population is grouped into three classes: susceptible, infectious, recovered individuals such that when susceptible (healthy) comes in contact of biological agents, it becomes infectious, and when infectious becomes healthy it comes into being recovered. Authors considered the homogeneously distributed and randomly mixed population and derived explicitly the notion of threshold parameter by introducing the basic reproduction number (R0). A reproduction number defines the average value of secondary infections by only one infectious individual during its infectious period. Kermack et. al. Computed that if R0<1, then infective decreases, whereas if R0<1, then the infective increases. This 
model is global considering topological features and the contact structures between different compartments separately. Apart from global models, there are also models where population is classified into some compartments taking into consideration of both topology structure and worm interactions features. For example, infectious class can also divided into infectious individual with 1 neighbor (1-degree infective), infections of 2 neighbors (2 degree infective), etc. K degree distribution is defined when individual keeps in contact of k neighbors. Based on these paradigms, models are investigated in different environments~\cite{ref17,ref728}.
 
Deterministic epidemic models derive the set of differential equations and their solutions~\cite{ref7a,ref7b,ref1a}. The system stability of the differential equations is also determined mathematically to measure the equilibrium conditions~\cite{ref7b,ref12,ref1a,ref10,ref11}. Based on different structure of a network, some models was proposed using sleep-wakeup mode~\cite{ref9a}, clustering \& community structure~\cite{ref19,ref22}, energy-drain, collisions \& congestions etc~\cite{ref25,ref26}. Majority of work are related to basic SI model and its modifications. Similarly, the global stability analysis are investigated in~\cite{ref16,ref18,ref24} based on the position for dysfunctional and based on IoT worm's attack . In these models, global behavior is captured to study the worm propagation dynamic with respect to total number of distributed nodes as mixed population (i.e., every node interacts with every one for infections). It does not take the account of local interactions of nodes, node's resources 
and data traffic influences, etc~\cite{ref12,ref13,ref14}.

\subsection{Spatial Correlation in WSN}~\label{Lsp}
\begin{figure}[t]
\centering
   \includegraphics[trim=0.0cm 0cm 0cm 0.7cm, clip=false,
scale=0.40]{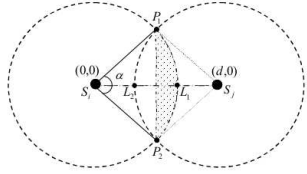}
 \caption{Spatial correlation between nodes $n_i$ and $n_j$ at locations $s_i$ and $s_j$~\cite{ref5}. }
 \label{fig:Net1}
\end{figure}

In our previous work in~\cite{ref3,ref4,ref5}, it can be shown that if control center (i.e. the sink) knows the locations of nodes and sensing range $r_s$, it can estimate the correlation using correlation coefficient between nodes. It gives the relationship between correlation coefficient between measured sensory data, $\rho_{(i,j)}$ and distance, $d_{(i,j)}$ between nodes $n_i$ and $n_j$ . Following the our works and Fig.~\ref{fig:Net1}, the expression is given by Eq.~\eqref{eq:corr1}.  

 \begin{table}
  \begin{center}
  \caption{Notations Summary}
 \begin{tabular}{||c|p{6.1cm}||} \hline
 \emph{Notation} & \emph{Description}  \\ \hline
 {$t$}& Time \\ \hline
 {$S(t)$}& Susceptible nodes at $t$ time \\ \hline
 {$I(t)$}& Infectious nodes at $t$ time \\ \hline
 {$R(t)$}& Recovered nodes at $t$ time \\ \hline
 {$d_{(i,j)}$}&{Euclidean distance between nodes $n_i$ and
 $n_j$} \\ \hline
 {$r_s$ }&{Sensing range of a node} \\ \hline
 {$r_t$ }&{Transmission range of a node} \\ \hline
 {$\rho _{(i,j)}$}& Correlation coefficient between  nodes $n_i$ and
 $n_j$ located at coordinates $s_i$ and $s_j$ \\ \hline
 {$\vartheta$}& Control parameter a variable to control the degree of correlation between nodes ($\vartheta = 2r_s$)\\ \hline
 {$N$}& Total number of nodes in the network \\ \hline
 {$\Delta t$}& A very short period of time \\ \hline
 {$r_e$}& Radius of occurring event in the network \\ \hline
  {$r_e(t)$}& Radius of infection spread at $t$ time \\ \hline
 \end{tabular}
 \label{tab123}
 \end {center}
 \end{table}

\begin{equation}
\begin{split}
\label{eq:corr1} 
\rho_{(i,j)}=K_{\vartheta}(d_v)&=\begin{matrix}
 \frac{\cos^{-1}{(\frac{d_v{(i,j)}}{\vartheta})}}{\pi} -
\frac{d_v{(i,j)}}{\pi\vartheta^2}.\sqrt{(\vartheta^2-d_v^2{(i,j)})}
\end{matrix} \\ &\begin{array}{ll} &\qquad \qquad \qquad \qquad \mbox{ For $0
\leq d_v{(i,j)}< \vartheta;$} \end{array} \\
&=0    \qquad  \quad\;  \; \,   \qquad \qquad \mbox{ For $d_v{(i,j)} \geq
\vartheta$}.
\end{split}
\end{equation}

\begin{figure}
\begin{center}
\begin{tabular}{cc}
\hspace{-0.020in}
\scalebox{.70}{\includegraphics[trim=0.5cm 0cm 0.7cm 0cm, clip=false,
scale=0.80]{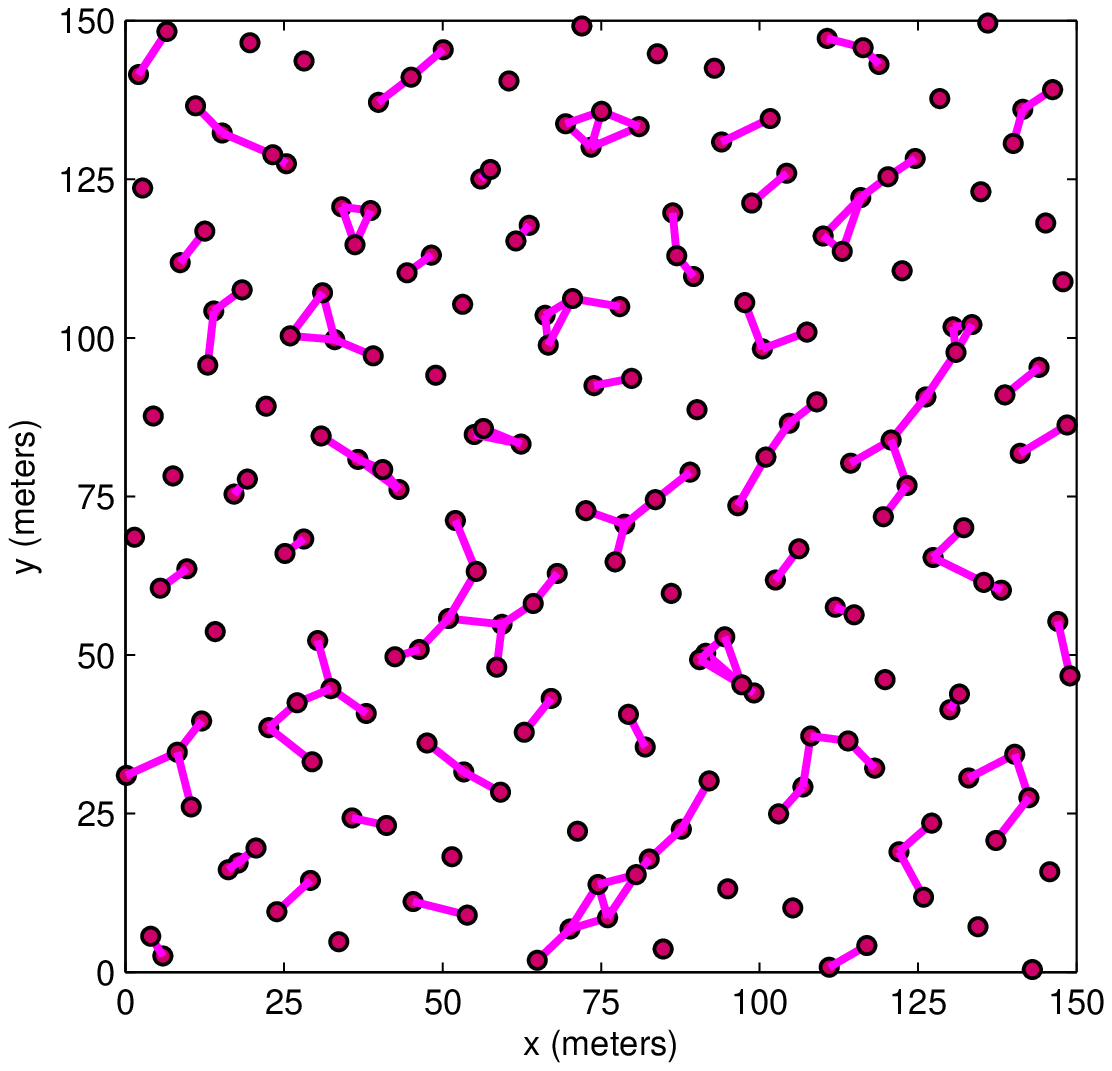}} & 
\scalebox{.70}{\includegraphics[trim=2.4cm 0cm 0cm 0cm, clip=false,
scale=0.80]{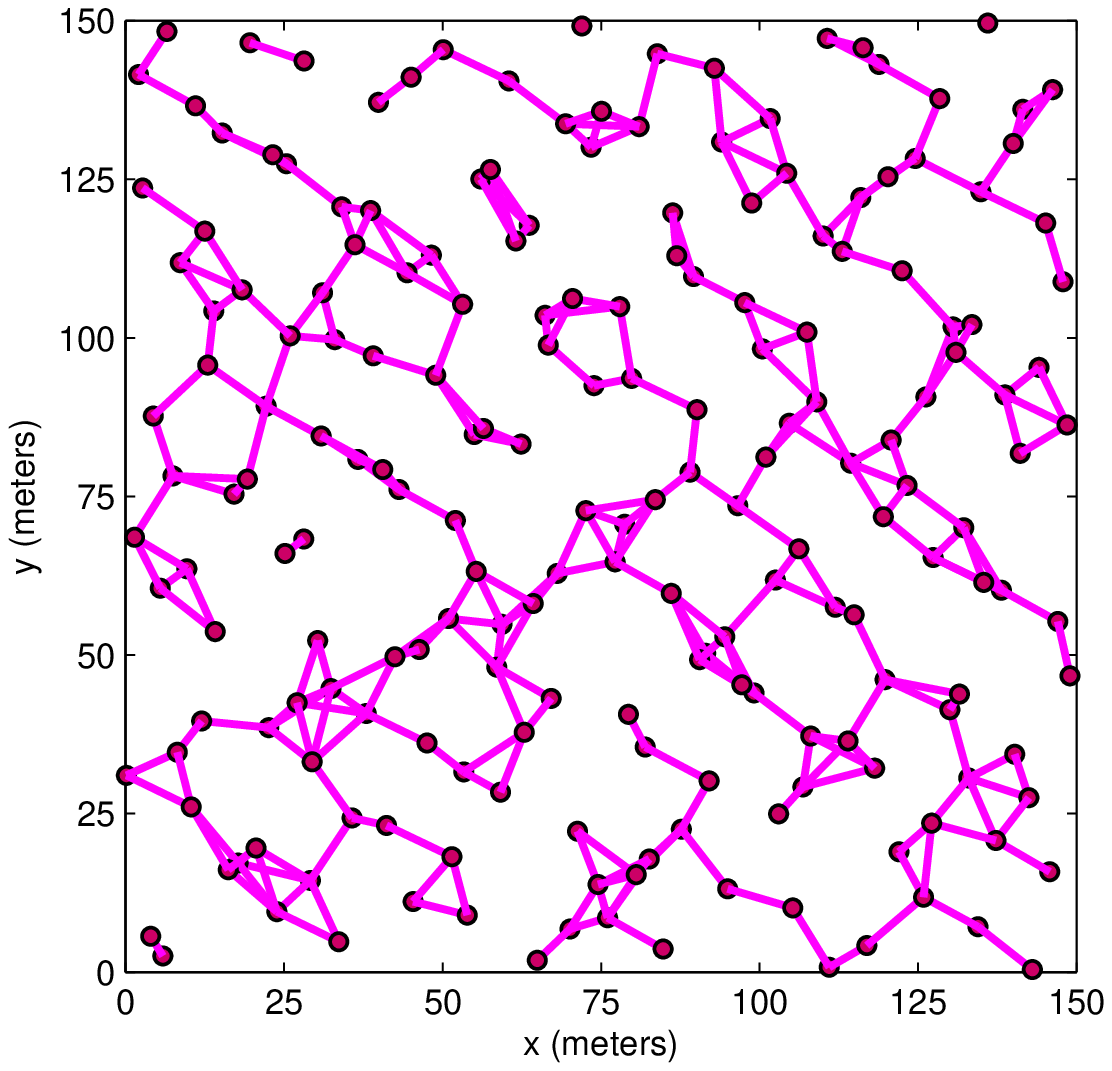}} \\
{\hspace{0.280in}\bf (a)} &\hspace{-0.520in}{\bf (b)}\\
\end{tabular}
 \hspace{3in}\parbox{5.5in}{\caption{Random distribution of 200 nodes with
(a) $\theta_1 = 9\;m$ and (b) $\theta_2 = 12\;m$.\label{fig:Net2}}}
\end{center}
\end{figure}

As per Eq.~\eqref{eq:corr1}, a subregion is a set of points that is formed by overlapped
sensing region between nodes of $r_s$-radius disk centered at position of itself
such that two spatial points belong to the same subregion if and only if they
are in overlapped coverage area of same set of sensor nodes. Thus, the spatial correlation between nodes is expressed by fraction of overlapped sensing area of nodes of $r_s$-radius disk centered at position of itself. So, $\rho_{(i,j)}$ gives the fraction of overlapped sensing area of nodes of $r_s$-radius disk with $d_{(i,j)}$ separation distance between them. Let us consider an example to get more insight in spatial correlation between nodes. We can have set of sensing ranges being $r=(2,\; 4.5,\;6,\;7.5,\;9,\;10)$, the set of control parameters $\vartheta$, can be $(\theta_1=4, \theta_2=9,\theta_3=12,\theta_4=15,\theta_5=18,\theta_6=20)$ respectively for different sensors attached with a node. We simulate $200$ randomly distributed nodes in a $150\times 150\;m^2$ area as shown in Figs.~\ref{fig:Net2}(a) and \ref{fig:Net2}(b). The
correlation between nodes is portrayed with variations in the control parameter
$\vartheta$. If the value of $\rho_{(i,j)}$ between two nodes is greater than
zero, then they are shown using a connected solid line. In
Figs.~\ref{fig:Net2}(a) and \ref{fig:Net2}(b), when a node-pair does not show
any correlation ($\rho_{(i,j)}$ is equal to zero) both the nodes are out
of sensing range from each other and there is no connecting line between them.
Fig~\ref{fig:Net2}(a) with a distribution of 200 nodes with $\theta_1 = 9\;m$
(or $r = 4.5\;$m) shows only a few connected lines indicating that very few
nodes are correlated with their neighbors. Given the same location of the
nodes, if we change the sensing 
range to $\theta_2 = 12\;m$ (Fig.~\ref{fig:Net2}(b)), more connected lines
appear. This indicates that more nodes are now correlated with their
neighboring
nodes. When node density changes for a given fixed sensing range (i.e.
$\theta_1$), more neighboring nodes are correlated (Figs.
\ref{fig:Net2}(b) and~\ref{fig:Net22}(a)). Fig.~\ref{fig:Net22}(b) shows the
correlation values on the lines connecting two nodes.

\begin{table}
\begin{center}
\caption{Sample results of $w_{\theta}$ for different values of $\theta$ }
\begin{tabular}{|c|*{5}{c|}}\hline
 Random Topology& $\theta=0.4$ &$\theta=0.4$&$\theta=0.5$ &$\theta=0.6$&$\theta=0.8$\\\hline
 $N\times N$& $\theta=0.4$ &$\theta=0.4$&$\theta=0.5$ &$\theta=0.6$&$\theta=0.8$\\\hline
 \end{tabular}
\label{tab:2}
\end{center}
\end{table}
In this way, the spatial correlation between nodes play role in redundant data transmission. The multiple messages about an event are transmitted to the sink due to common overlapped sensing. To investigate the impact of this in virus spread behavior if the nodes are attacked by malware, we define correlation weight denoted by $w_i$ of a node $i$ using $\rho_{(i,j)}$. 
\begin{equation}
\label{eqW1}
 w_i = \sum \limits_{i=1}^{n}{\sum \limits_{j=1,j\neq1}^{n}{\rho_{(i,j)}}}
\end{equation}
It can be seen from above Eq.~\eqref{eqW1} that each node has correlated connection with nearby nodes depending on overlapped sensing value. In particular, if we define $w_{\theta}$ as the average degree of correlation in a network for given $\theta$. Then the spread of infection can be captured with $w_{\theta}$ for connected neighbors of the nodes based on sensing range and communication range. Using this parameter, the behavior of virus dynamics with spatial correlation is studied that takes into account of strongly correlated nodes and weakly correlated nodes with respect to random node distribution. Using this realistic parameters, a modified SI model is derived mathematically and its performance comparisons is discussed in details. In next section, existing SI model of worm propagation is discussed in short. Then, new modified SI model is described.

\section{SI Epidemic Models for worm propagation}

In this section, a short overview of epidemic SI models for worm propagation in WSNs is presented before our model design. The sleep-wakeup mode of sensor nodes has been used by Tang~\cite{ref9a} for modeling SI for WSNs. Author has used the duty cycle of nodes to cure the infective nodes in improved version of susceptible-infectious (SI) model. 

\subsection{SI Model Without Anti-Virus Mechanism}
By Tang~\cite{ref9a}, followings are the expressions for SI model without anti-Virus Mechanism in sleep mode. If $\beta$ is rate of infection possibility to become an infective through contact, then  the SI model is given without anti-virus mechanism as

 \begin{equation}
 \begin{split}
\label{SI0}
 	\frac{dI(t)}{dt} &= -2\beta(\sqrt{\sigma\pi}r_t)^3 \sqrt{I(t)} \frac{N-I(t)}{N},\\
 	\frac{dS(t)}{dt} &=  2\beta (\sqrt{\sigma \pi}r_t)^3 \sqrt{I(t)} \frac{N-I(t)}{N}.
\end{split}
\end{equation}
The solution of above equations for $I(t)$ is given as
\begin{equation}
 \label{SIZ1}
  	I(t) = \left(\begin{array}{c} \frac{2}{1+\frac{\sqrt{N}-1}{\sqrt{N}+1}e^{-\frac{2\beta(\sqrt{\sigma\pi}r_t)^3t}{\sqrt{N}}}}-1\end{array}\right)^2.
\end{equation}
Where $S(t) + I(t) = N$. For uniformly distributed $N$ nodes with node density $\sigma$, each infective node do not take part in the infection of other susceptible nodes. But the infective nodes having susceptible nodes are responsible to spread infection at a time.
\subsection{SI Model With Anti-Virus Mechanism in Sleep Mode}
Tang~\cite{ref9a} utilizes the sleep-wakeup duty cycle feature to detect the infection. A node can detect the its state of infectious by running anti-virus mechanism in sleep mode. In WSNs, most of time, there is no activity, so nodes undergo sleep mode for power saving. If a node gets infected, then it can cure itself during sleep mode by running anti-virus mechanism. Let $p$ be the fraction of maintained infective nodes to be cured within predefined time period and $\lambda$ be the rate of maintenance (i.e., duty cycle). The $SI$ epidemic with anti-virus mechanism is given as
 \begin{equation}
 \begin{split}
\label{SIa1}
 	\frac{dI(t)}{dt} &= -2\beta(\sqrt{\sigma\pi}r_t)^3 \sqrt{I(t)} \frac{N-I(t)}{N}+\lambda pI(t),\\
 	\frac{dS(t)}{dt} &=  2\beta (\sqrt{\sigma \pi}r_t)^3 \sqrt{I(t)} \frac{N-I(t)}{N}-\lambda pI(t).
\end{split}
\end{equation}
The solution of above equations for $I(t)$ is given as
\begin{equation}
\begin{split}
\label{SIa2} 
I(t)&=\begin{matrix}
{\left(\begin{array}{c}\frac{B+C}{1+\frac{C-1}{B+1}exp(-\frac{1}{2}A(B+C)t)}-B\end{array}\right)}^2
\end{matrix} \\ &\begin{array}{ll} &\qquad \qquad \qquad \qquad \mbox{For $I(t)
\leq C^2;$} \end{array} \\
&=\begin{matrix}
{\left(\begin{array}{c}\frac{B+C}{1-\frac{C-1}{B+1}exp(-\frac{1}{2}A(B+C)t)}-B\end{array}\right)}^2
\end{matrix} \\ &\begin{array}{ll} &\qquad \qquad \qquad \qquad \mbox{For $I(t)
\geq C^2.$} \end{array}
\end{split}
\end{equation}
Where, \begin{equation}A = \frac{2\beta}{N} (\sqrt{\sigma\pi}r_t)^3, \\\nonumber
       \end{equation} \begin{equation}B = \sqrt{{(\frac{\lambda p}{2A})}^2+N}+\frac{\lambda p}{2A}, \\\nonumber \end{equation} \begin{equation}C = \sqrt{{(\frac{\lambda p}{2A})}^2+N}-\frac{\lambda p}{2A}.\\\nonumber \end{equation}

\section{SI Model with Spatial Correlation }
In the section~\ref{Lsp}, we show that how spatial correlation exist in WSN based on node density and position of nodes. Each node has specific correlation degree denoted by $w_{\theta}$. This is taken into account to investigate the dynamics in WSNs. Assuming total $N$ sensor nodes are deployed with $\sigma$ node density in a monitored area. Each sensor node has $r_s$ sensing range and $r_t$ transmission range (here $r_t \geq 2r_s$). There are two set of nodes namely Susceptible node set ($S$), Infectious node set ($I$). The $\beta$ is rate of infection possibility to become an infective. For event-driven sensor system, correlated nodes are usually observed based on $r_s$ sensing range. Since these nodes are highly responsible for virus propagation in WSN, we take into the account using $w_{\theta}$. We get
\begin{equation}
\label{SI1}
 	\frac{dI(t)}{dt} = \beta \sigma \pi r^2_t I_{e}(t)S(t)w_{\theta}.
\end{equation} 
\begin{equation}
\label{SI2}
 	\frac{dS(t)}{dt} = - \beta \sigma \pi r^2_t I_{e}(t)S(t)w_{\theta}.
\end{equation} 

\begin{figure}[!h]
\begin{center}
\begin{tabular}{c}
\hspace{-0.35in}
\includegraphics[trim=0.0cm 0cm 0cm 0.7cm, clip=false,
scale=0.25]{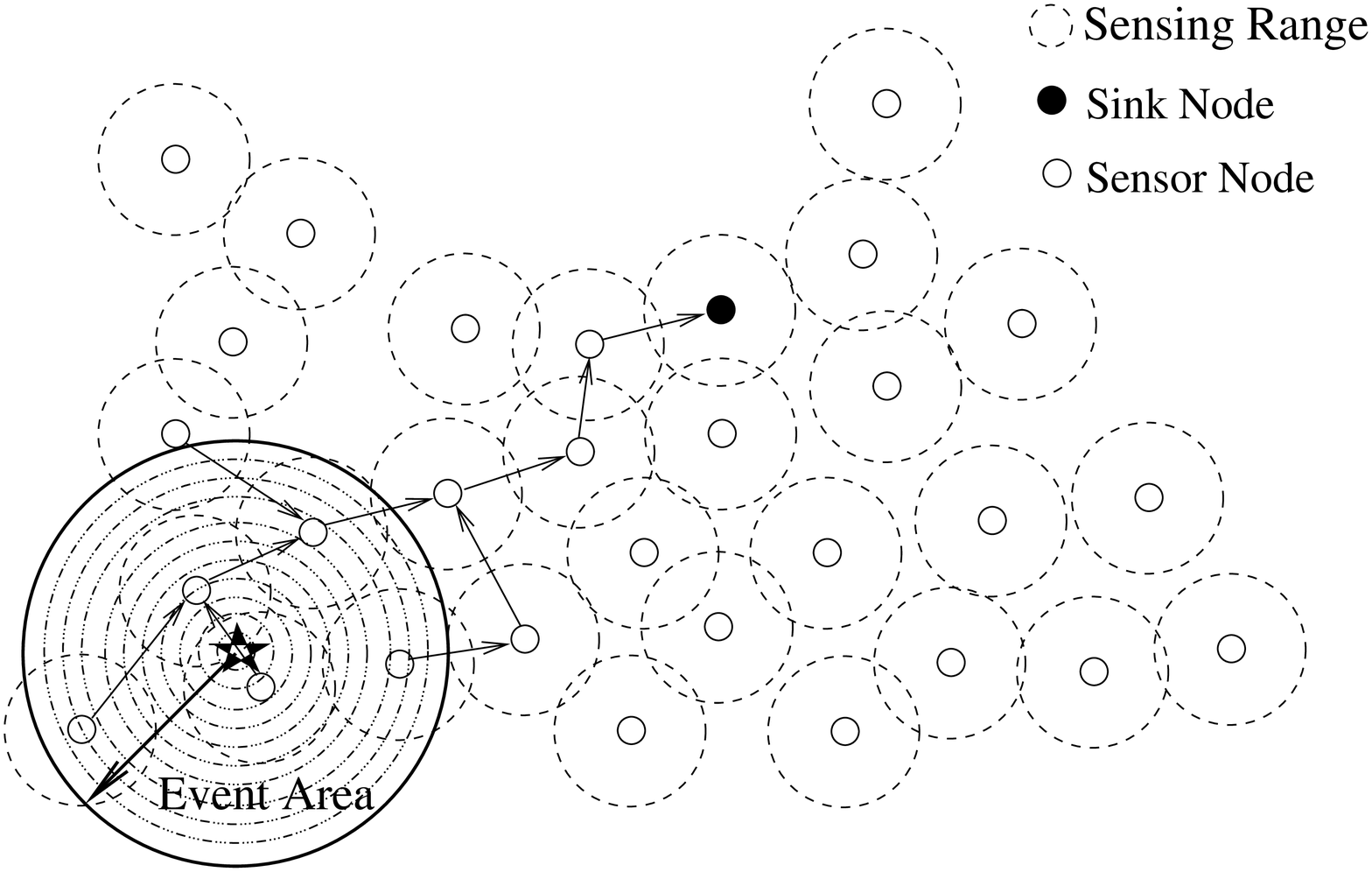} \\
\includegraphics[scale=0.70]{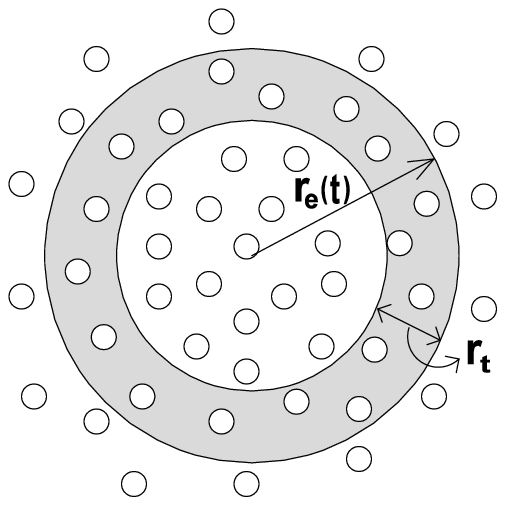}\\ 
\end{tabular}
 \hspace{3in}\parbox{4.8in}{\caption{ Typical sensor network with $r_s$ sensing range and $r_e$-radius event area and the effective shaded circular strip for Virus Spreading (Right side). \label{fig1}}}
\end{center}
\end{figure}
%
Assuming the fluctuating number of sensor nodes is from time $t$ to $\Delta t$ where $\Delta t \geq 0$ and $\Delta t$ denotes very short time period. According to mean-field theory, the valid fluctuating number of infectious sensor nodes at $\Delta t$ time is given by 
\begin{equation}
\label{sird1}
 	I(t+\Delta t) = (\beta \sigma \pi r^2_t I_{e}(t)S(t)w_{\theta})\Delta t.
\end{equation} 
\begin{equation}
 \label{sird14}
 	S(t) + I(t) = N.
\end{equation}
Let a sensor node detects false event by means of virus attack. When a sensor node gets infection of viruses, then it becomes infected source node. The $r_e$-radius sized event area is formed by this false event as shown in Fig.~\ref{fig1}. So, all the susceptible nodes covered by $r_e$-radius sized event area would like to be infected source nodes. All these infectious nodes will not be responsible to infect others outside the event area. The spread of virus infection can only be possible by local interactions (i.e., communication within transmission range). Since the nodes have $r_t$ transmission range to communicate, the virus infection would only be from the effective infectious nodes that are inside the border area of $r_e$-radius circle (i.e., shaded circular strip shown in Fig.~\ref{fig1}). Following expressions hold.

\begin{equation}
\label{sird2}
 	i_e(t) = \frac{ \sigma  \pi r^2_e(t) -  \sigma \pi {(r_e(t)-r_t)}^2}{\sigma \pi r^2_t} \approx \frac{2 r_e(t)}{r_t}.
\end{equation} 	
As stated that the infection is introduced by infectious nodes around shaded circular strip area shown in Fig.~\ref{fig1}. At time $t$, $r_e(t)$ is the radius of infection spread and $I(t)$ is the number of infectious nodes that are responsible to infect the others. So, we get
\begin{equation}
\label{sird3a}
 	I(t) = \sigma \pi r_e^2(t).
\end{equation}
Using Eq.~\eqref{sird2} and Eq.~\eqref{sird3a}, we get
\begin{equation}
\label{sird3b}
 	i_e(t) = \frac{2}{ r_t}\sqrt{\frac{I(t)}{\sigma \pi}}.
\end{equation}
If there are total $N$ nodes considered, then we have
\begin{equation}
\label{sird3bc}
 	I_e(t) = \frac{2}{ r_t}\sqrt{\frac{I(t)}{N\sigma \pi}}.
\end{equation}
Now, using Eq.~\eqref{SI1} and Eq.~\eqref{sird3bc},
\begin{equation}
\label{sird5}
 	I(t+\Delta t) = \left(\begin{array}{c}2\beta r_t \sqrt{\sigma \pi} \sqrt{\frac{I(t)}{N}}\left(\begin{array}{c}N-I(t)\end{array}\right)w_{\theta}\end{array}\right)\Delta t.
\end{equation}
Similarly, we get,
\begin{equation}
\label{sird6}
	S(t+\Delta t) = \left(\begin{array}{c}-2\beta r_t \sqrt{\sigma \pi} \sqrt{\frac{I(t)}{N}}\left(\begin{array}{c}N-I(t)\end{array}\right)w_{\theta}\end{array}\right)\Delta t.
\end{equation}

Therefore, the following differential equations are derived from Eqs.~\eqref{sird5} and \eqref{sird6}.

\begin{equation}
\label{SI3}
 	\frac{dS(t)}{dt} = -2\beta r_t \sqrt{\sigma \pi} \sqrt{\frac{I(t)}{N}}\left(\begin{array}{c}N-I(t)\end{array}\right)w_{\theta}.
\end{equation} 
\begin{equation}
\label{SI4}
 	\frac{dI(t)}{dt} = 2\beta r_t \sqrt{\sigma \pi} \sqrt{\frac{I(t)}{N}}\left(\begin{array}{c}N-I(t)\end{array}\right)w_{\theta}.
\end{equation} 
Let the initial conditions be $S(0)=S_0=N-1$, $I(0)=I_0=1$. The above equation~\eqref{SI4} can be The solution of above equations for $I(t)$ is given as
\begin{equation}
 \label{my1}
  	I(t) = N\left(\begin{array}{c} \frac{2}{1-\frac{1-\sqrt{N}}{1-\sqrt{N}}e^{-\frac{2 w_{\theta}\beta r_t\sqrt{N\sigma\pi}t}{\sqrt{N}}}}-1\end{array}\right)^2.
\end{equation}

\section{Simulation Results and Analysis}
In this section, simulation results are presented to study the dynamic characteristics of virus propagation and impact of spatial correlation for WSNs. Using MATLAB, we simulate non-linear differential equations of our model. Results are produced by changing various parameters. Assuming the probabilities $\alpha_{1-6}$ be $0.04$, $0.0003$, $0.0006$, $0.17$, $0.6$, $0.008$, respectively. Initial values are given as: $R(0)\;=\;0$; $D(0)\;=\;0$; $I(0)\;=\;1$. Theoretical analysis has been done in different case studies with different parameters. These case studies are following. We have assumed all the parameters as dimensionless units.

\begin{figure}[!h]
\begin{center}
\begin{tabular}{cc}
\hspace{-0.35in}
\includegraphics[trim=0.0cm 0cm 0cm 0.7cm, clip=false,scale=0.5]{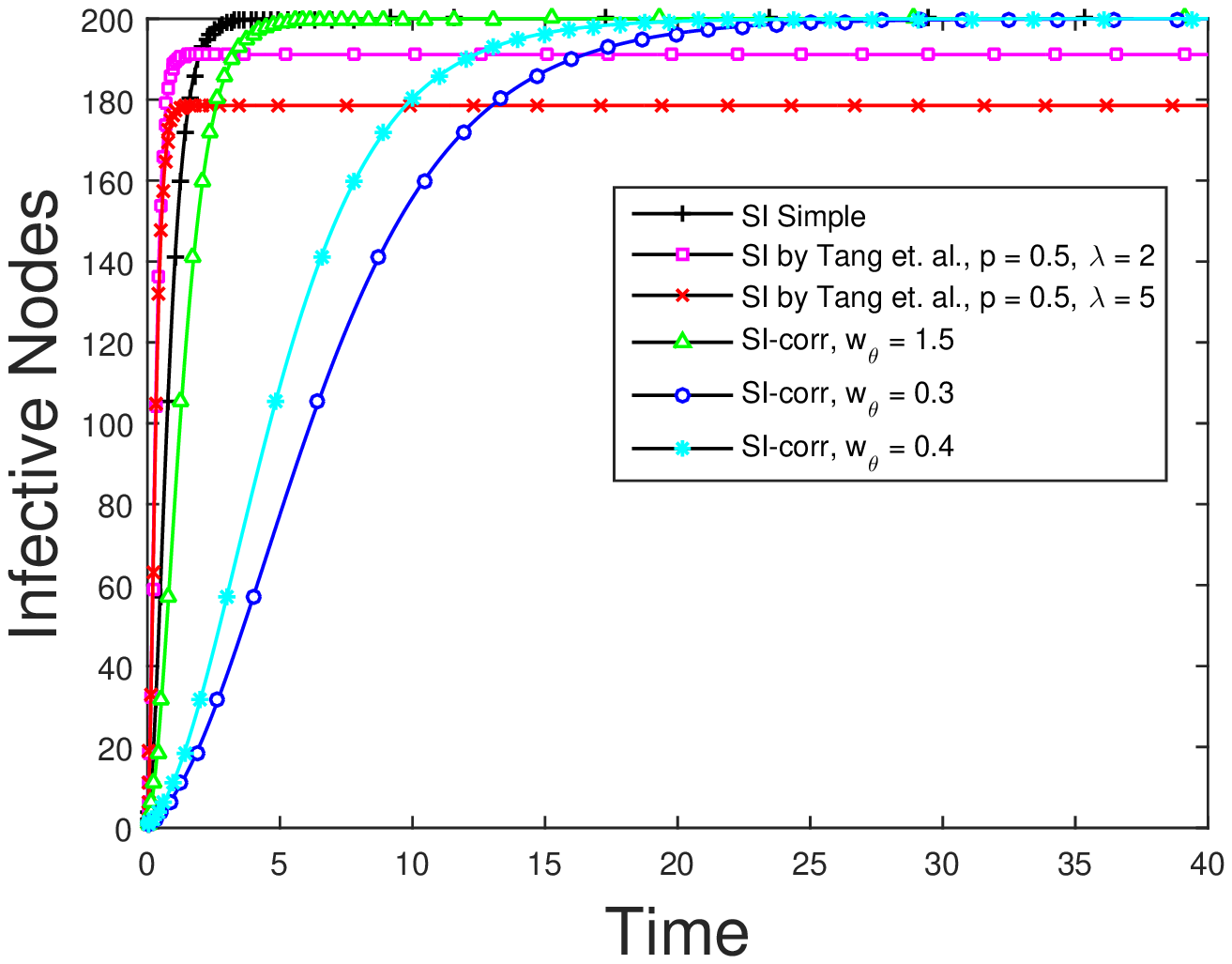} &
\includegraphics[scale=0.50]{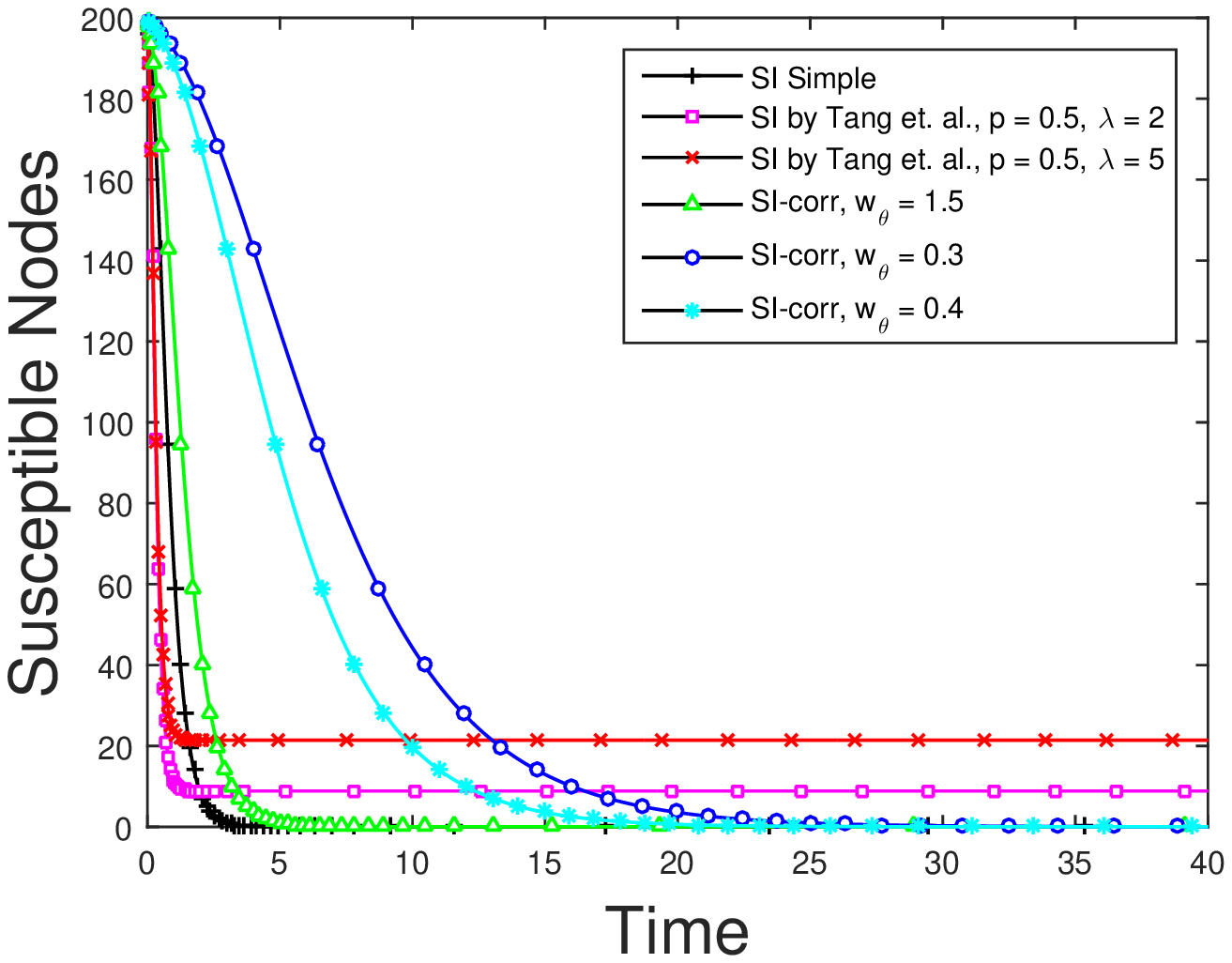}\\ 
\end{tabular}
 \hspace{3in}\parbox{4.8in}{\caption{Growth of infected nodes, I(t) (left figure) and susceptible nodes, S(t) (right figure) with time for different values of $w_{\theta}$ (spatial correlation degree) for constant values, $N=200$, $r_t=4$, $\sigma=0.5$, and $\beta=0.3$.Fig01}}
\end{center}
\end{figure}

In this group, when value of $\xi$ goes to lower with same node density (i.e., same $\lambda$), neighboring nodes are more due to highly correlated clusters formation as per correlation model. To evaluate the behavior of virus spread for this, the results using system Eq.~\eqref{sird13} are shown in Fig.~\ref{fig:Tra01}. We see the dynamics with time $t$ for $S(t)$, $I(t)$, $R(t)$, and $D(t)$ with four different values of correlation threshold $\xi$. In all the graphs, when number of infective nodes increases with time, susceptible nodes decreases. It is expected in the network because when more nodes get infected as they moves into infectious compartment. But with change in correlation in the network, there are changes in behavior of infectious and recovered compartments as shown in Fig.~\ref{fig:Tra01}.

\begin{figure}[!h]
\begin{center}
\begin{tabular}{cc}
\hspace{-0.35in}
\includegraphics[trim=0.0cm 0cm 0cm 0.7cm, clip=false,scale=0.5]{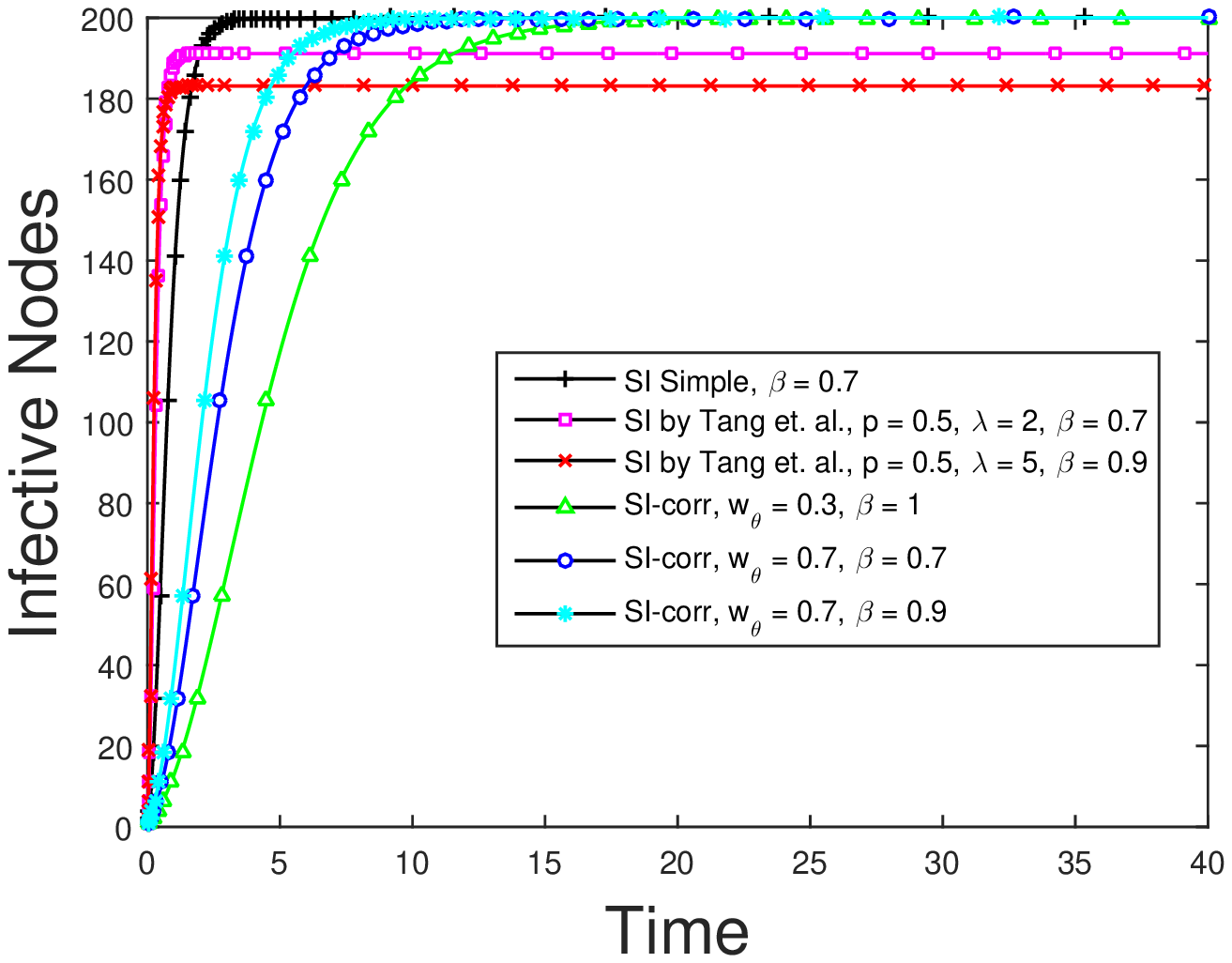} &
\includegraphics[scale=0.50]{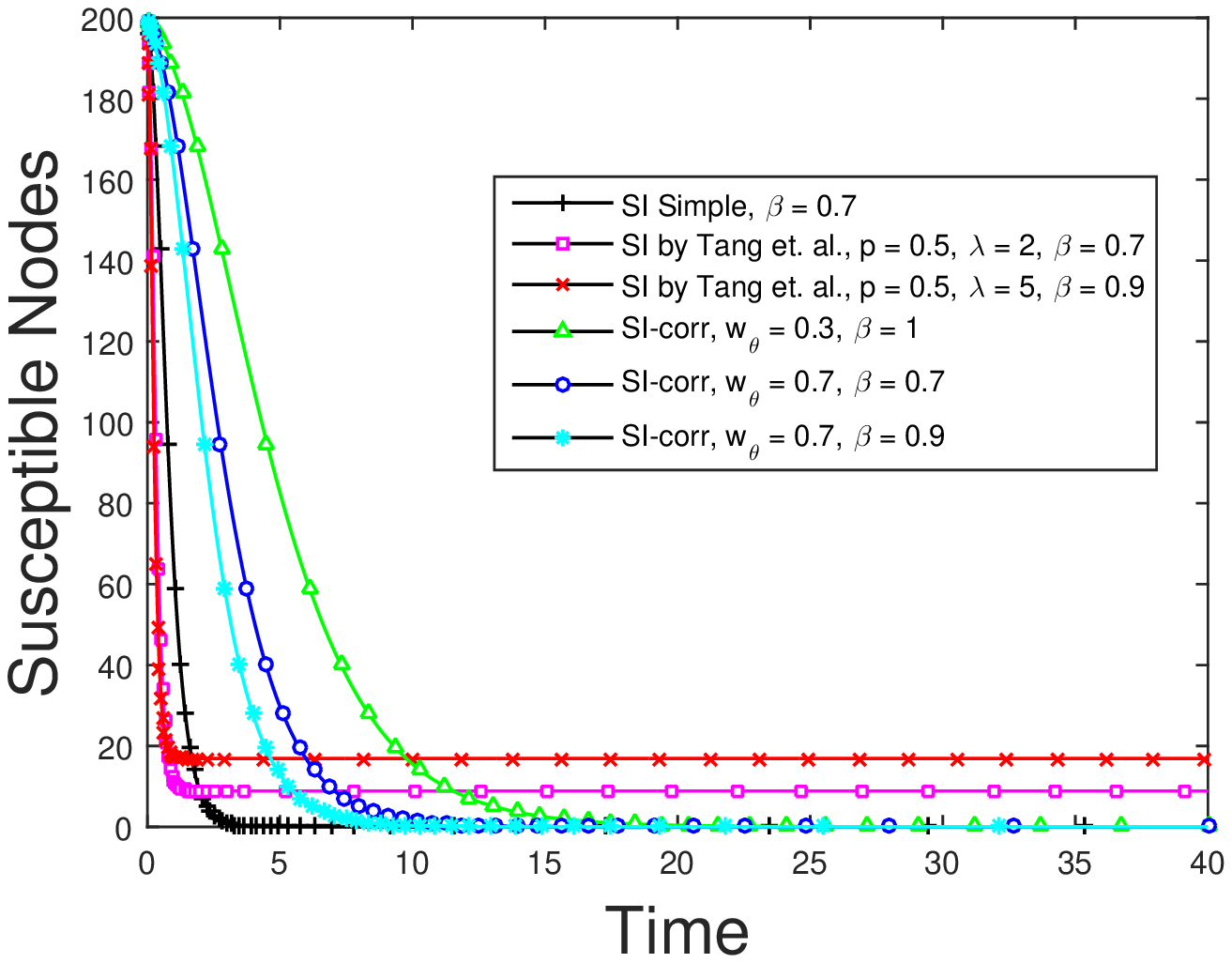}\\ 
\end{tabular}
 \hspace{3in}\parbox{4.8in}{\caption{Time dynamics of infected nodes, I(t) (left figure) and S(t) (right figure) with time for different values of $w_{\theta}$ (spatial correlation degree) and $\beta$ (infection capacity) for constant values, $N=200$, $r_t=4$, $\sigma=0.5$.}}
\end{center}
\end{figure}

\begin{figure}[!h]
\begin{center}
\begin{tabular}{cc}
\hspace{-0.35in}
\includegraphics[trim=0.0cm 0cm 0cm 0.7cm, clip=false,scale=0.5]{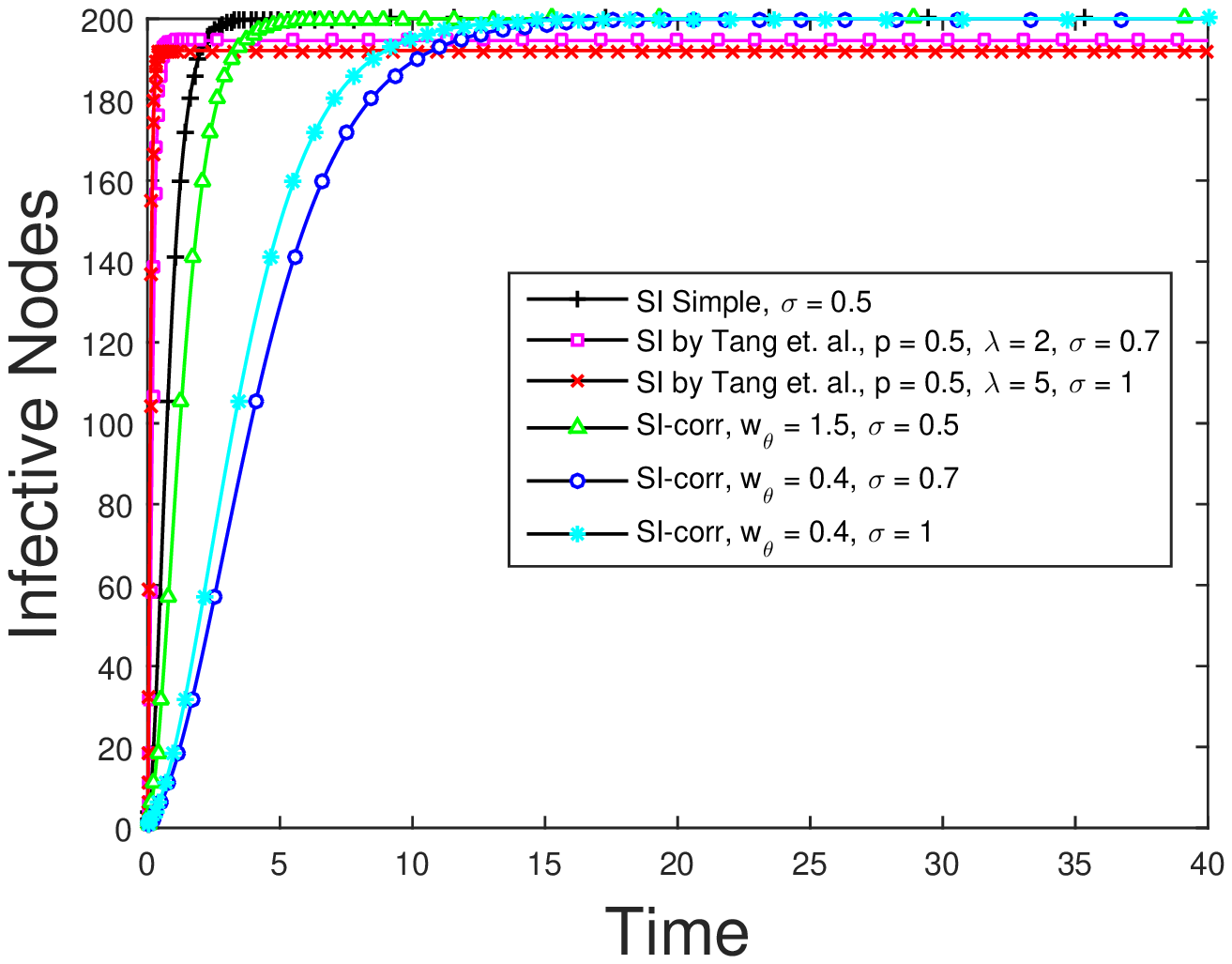} &
\includegraphics[scale=0.50]{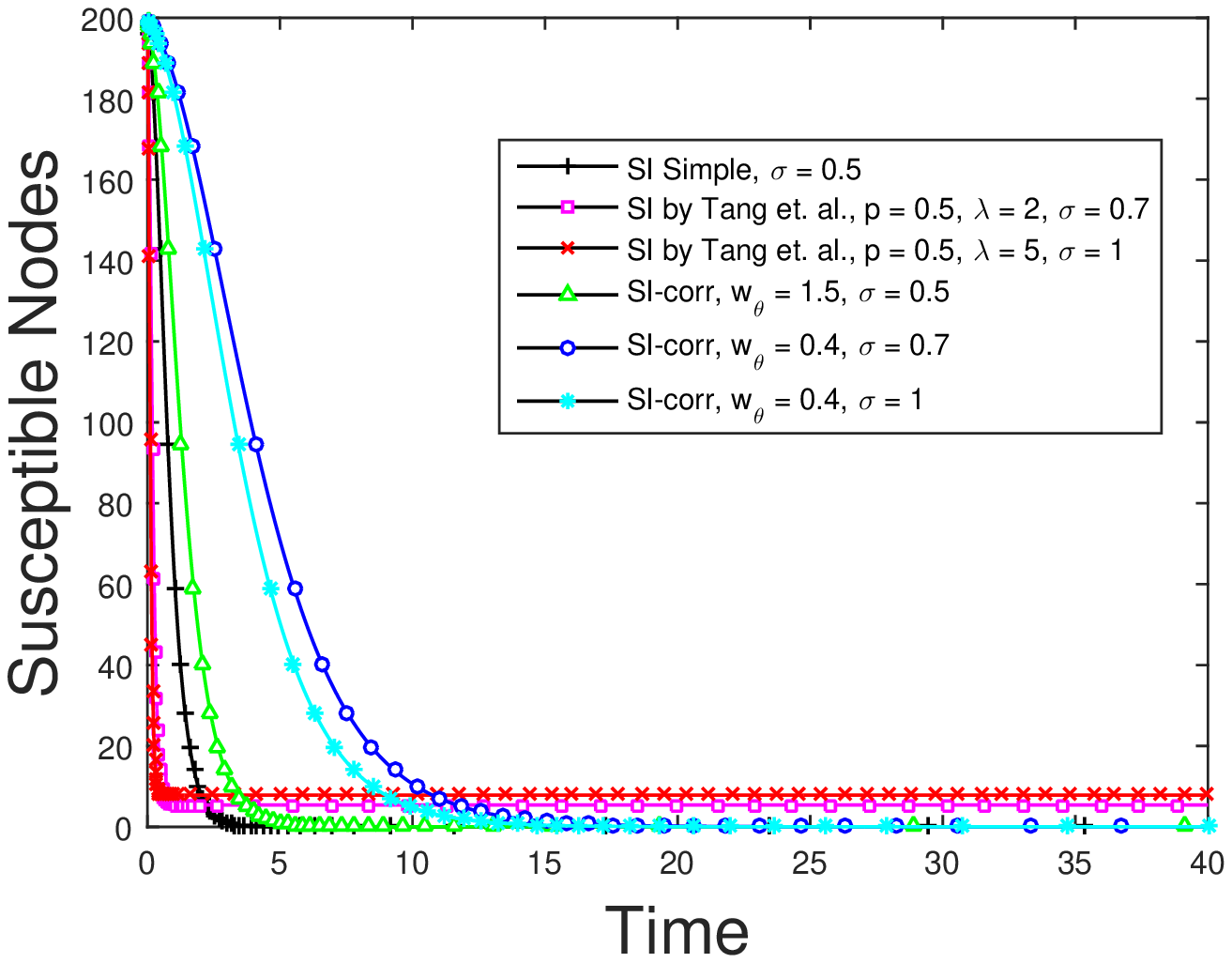}\\ 
\end{tabular}
 \hspace{3in}\parbox{4.8in}{\caption{Time dynamics of infected nodes, I(t) (left figure) and S(t) (right figure) with time for different values of $w_{\theta}$ (spatial correlation degree) and $\sigma$ (node density) for constant values, $N=200$, $r_t=4$, $\beta=0.3$.}}
\end{center}
\end{figure}

\begin{figure}[!h]
\begin{center}
\begin{tabular}{cc}
\hspace{-0.35in}
\includegraphics[trim=0.0cm 0cm 0cm 0.7cm, clip=false,scale=0.5]{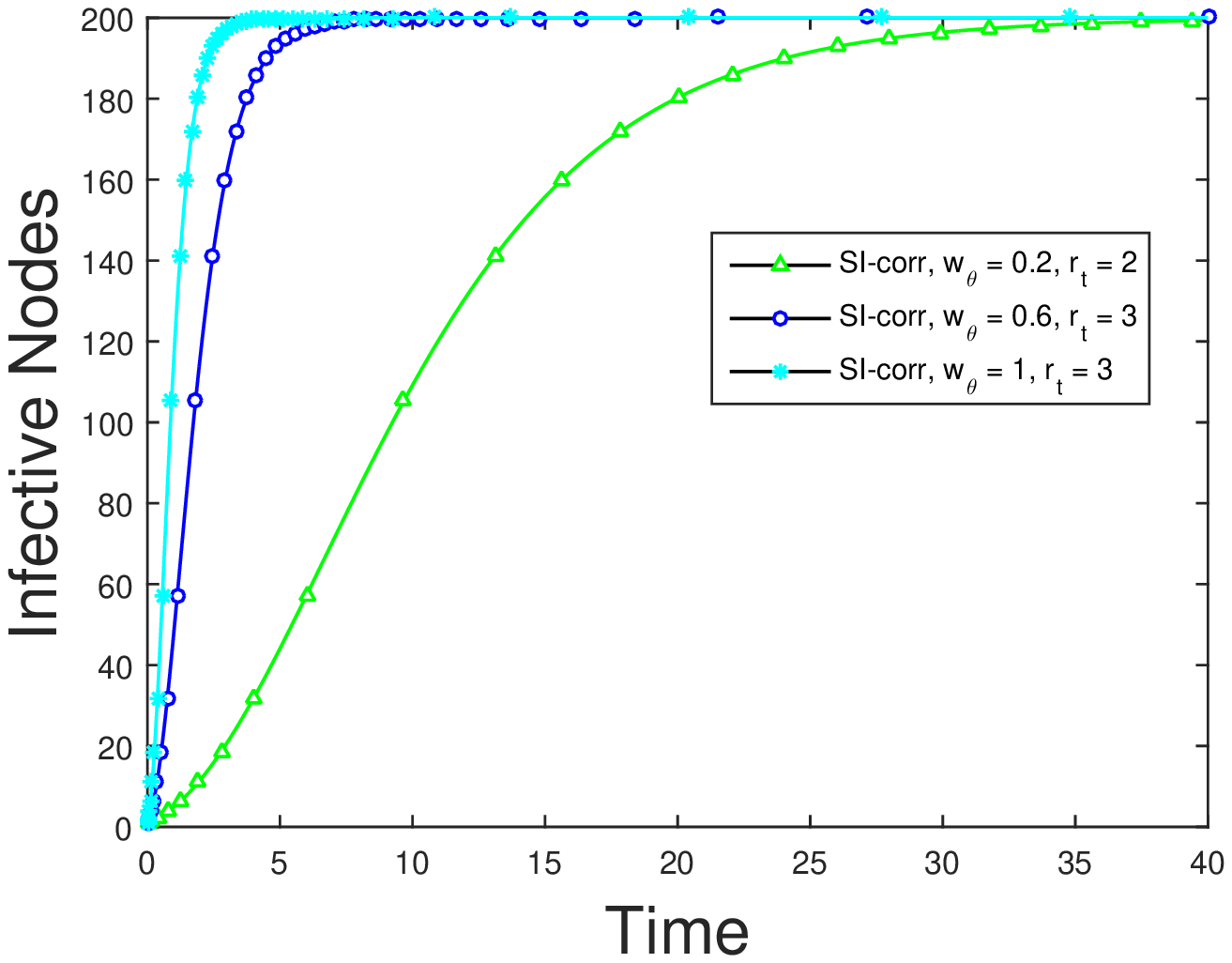} &
\includegraphics[scale=0.50]{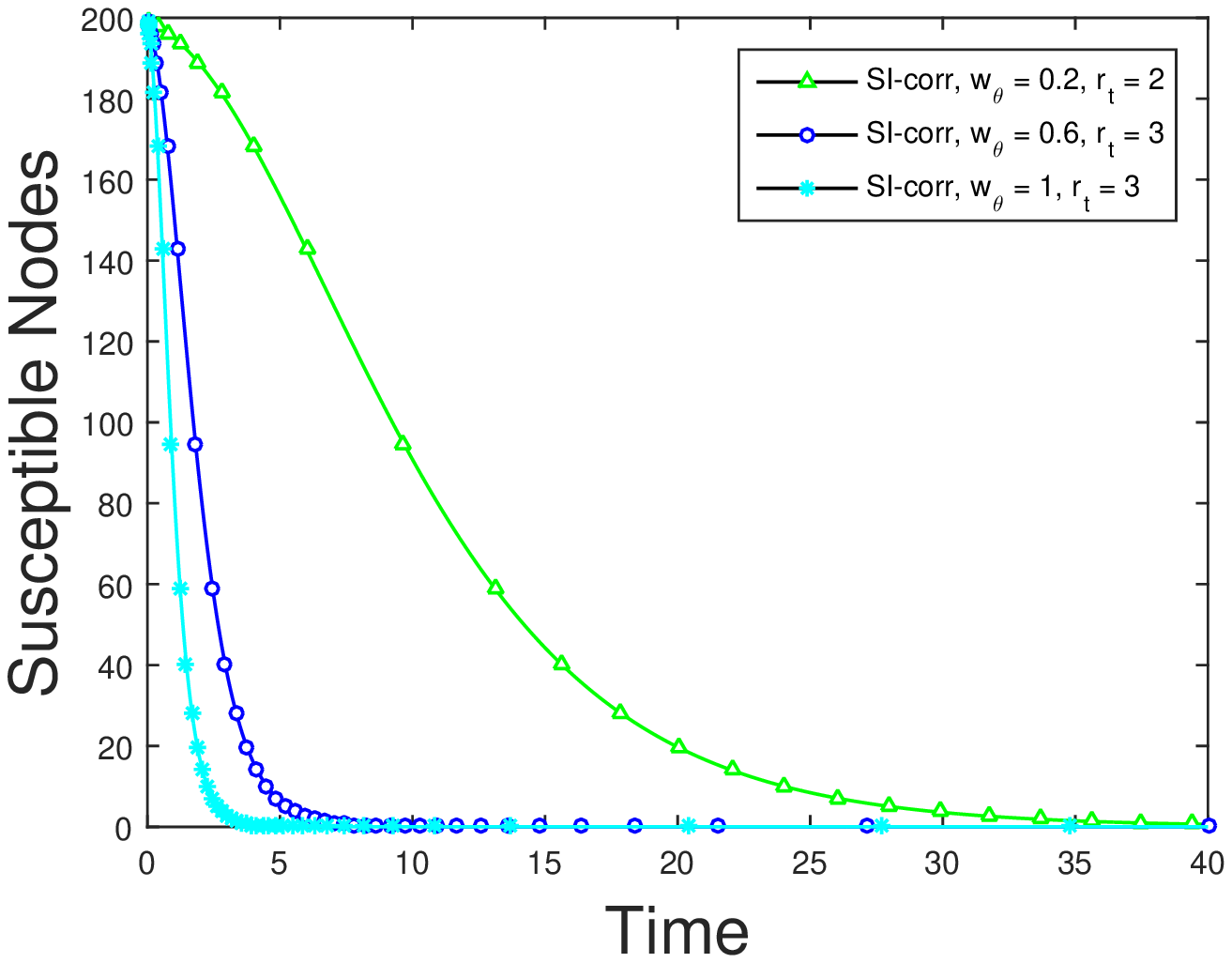}\\ 
\end{tabular}
 \hspace{3in}\parbox{4.8in}{\caption{Growth of infected nodes, I(t) (left figure) and S(t) (right figure) with time for different values of $w_{\theta}$ (spatial correlation degree) and $r_t$ (transmission range) for constant values, $N=200$, $\beta=0.3$, $\sigma=0.5$.}}
\end{center}
\end{figure}

\begin{figure}[!h]
\begin{center}
\begin{tabular}{cc}
\hspace{-0.35in}
\includegraphics[trim=0.0cm 0cm 0cm 0.7cm, clip=false,scale=0.5]{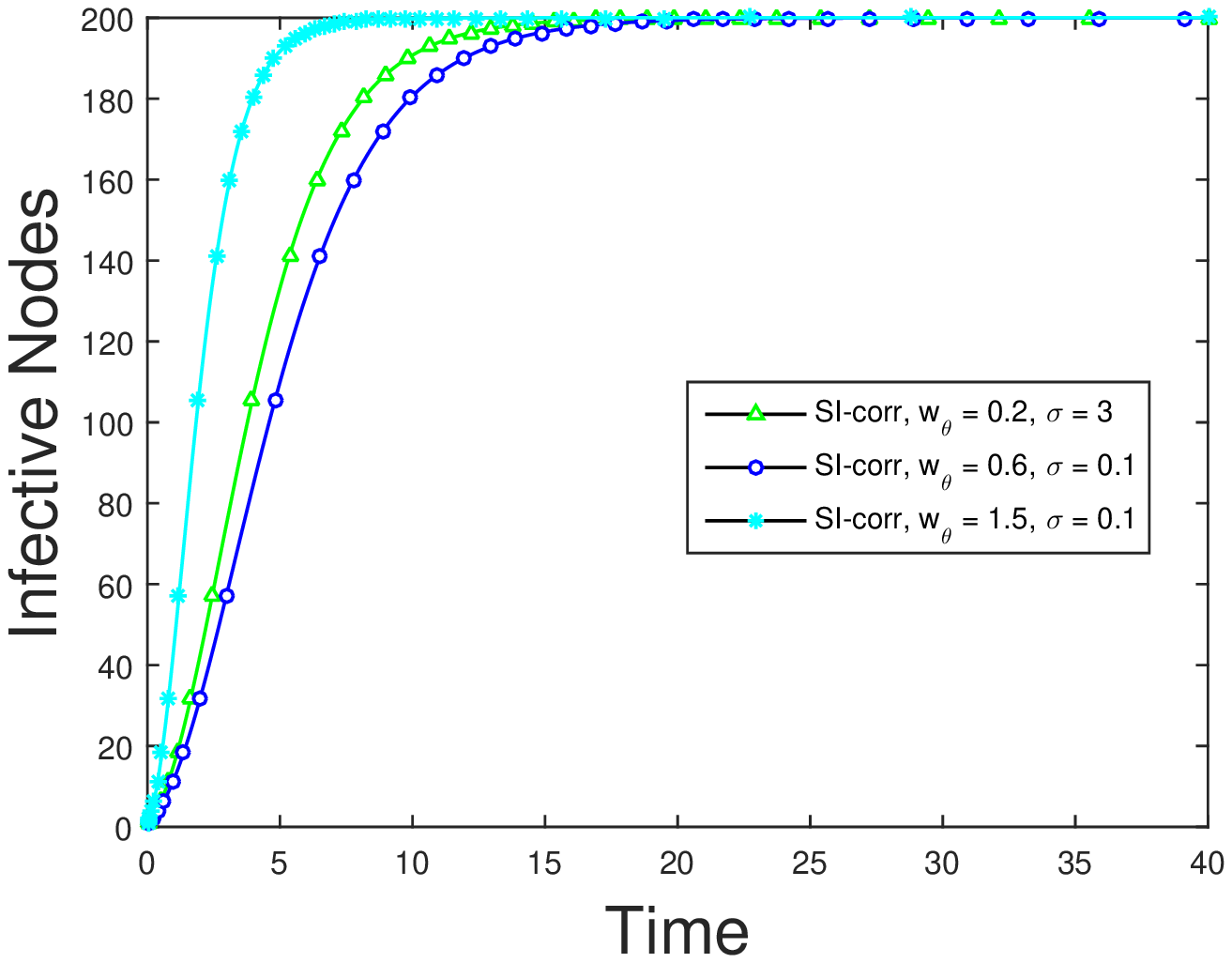} &
\includegraphics[scale=0.50]{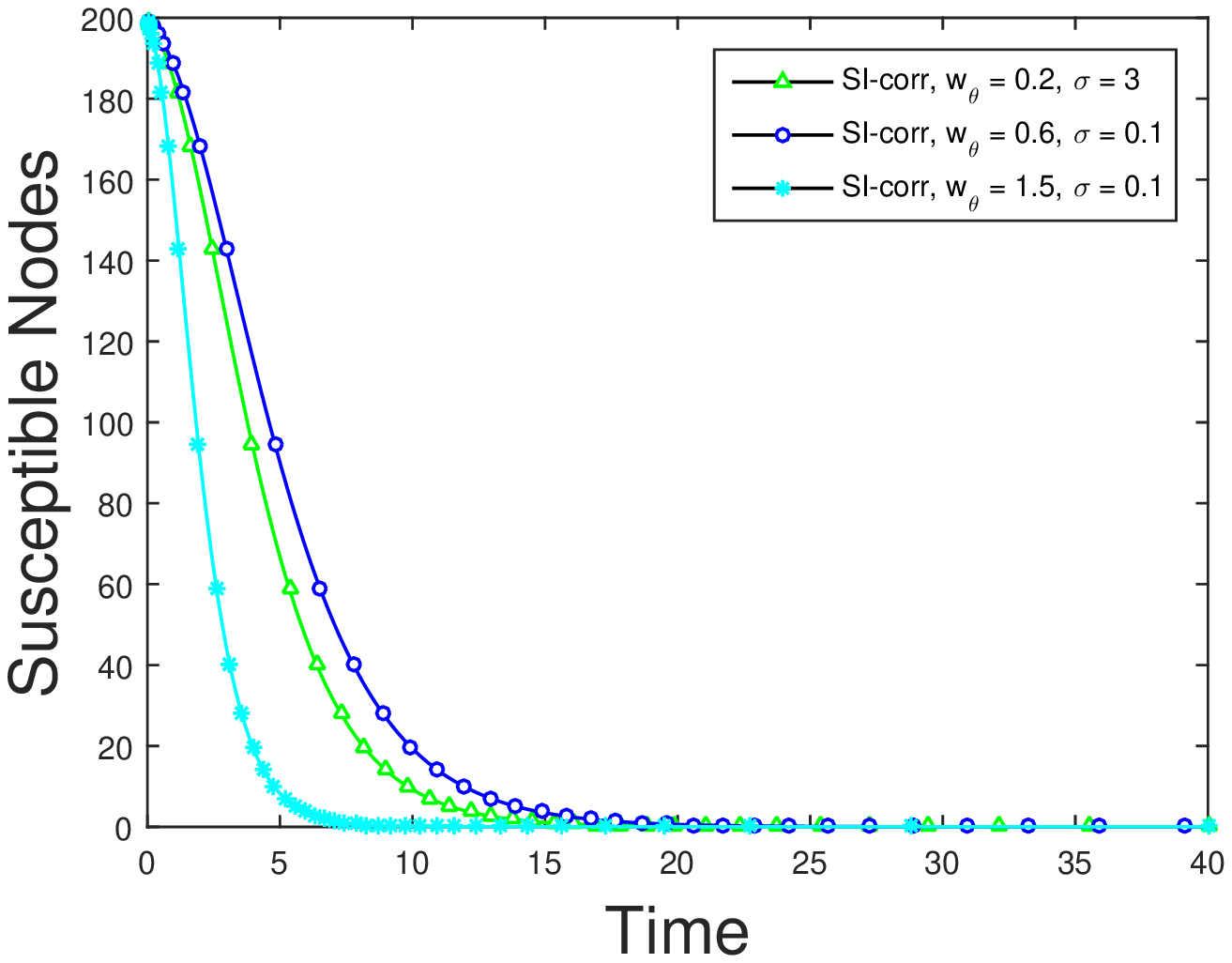}\\ 
\end{tabular}
 \hspace{3in}\parbox{4.8in}{\caption{Growth of infected nodes, I(t) (left figure) and S(t) (right figure) with time for different values of $w_{\theta}$ (spatial correlation degree) and $\sigma$ (transmission range) for constant values, $N=200$, $r_t=2$, $\beta=0.3$.}}
\end{center}
\end{figure}

\begin{figure}[!h]
\begin{center}
\begin{tabular}{cc}
\hspace{-0.35in}
\includegraphics[trim=0.0cm 0cm 0cm 0.7cm, clip=false,scale=0.5]{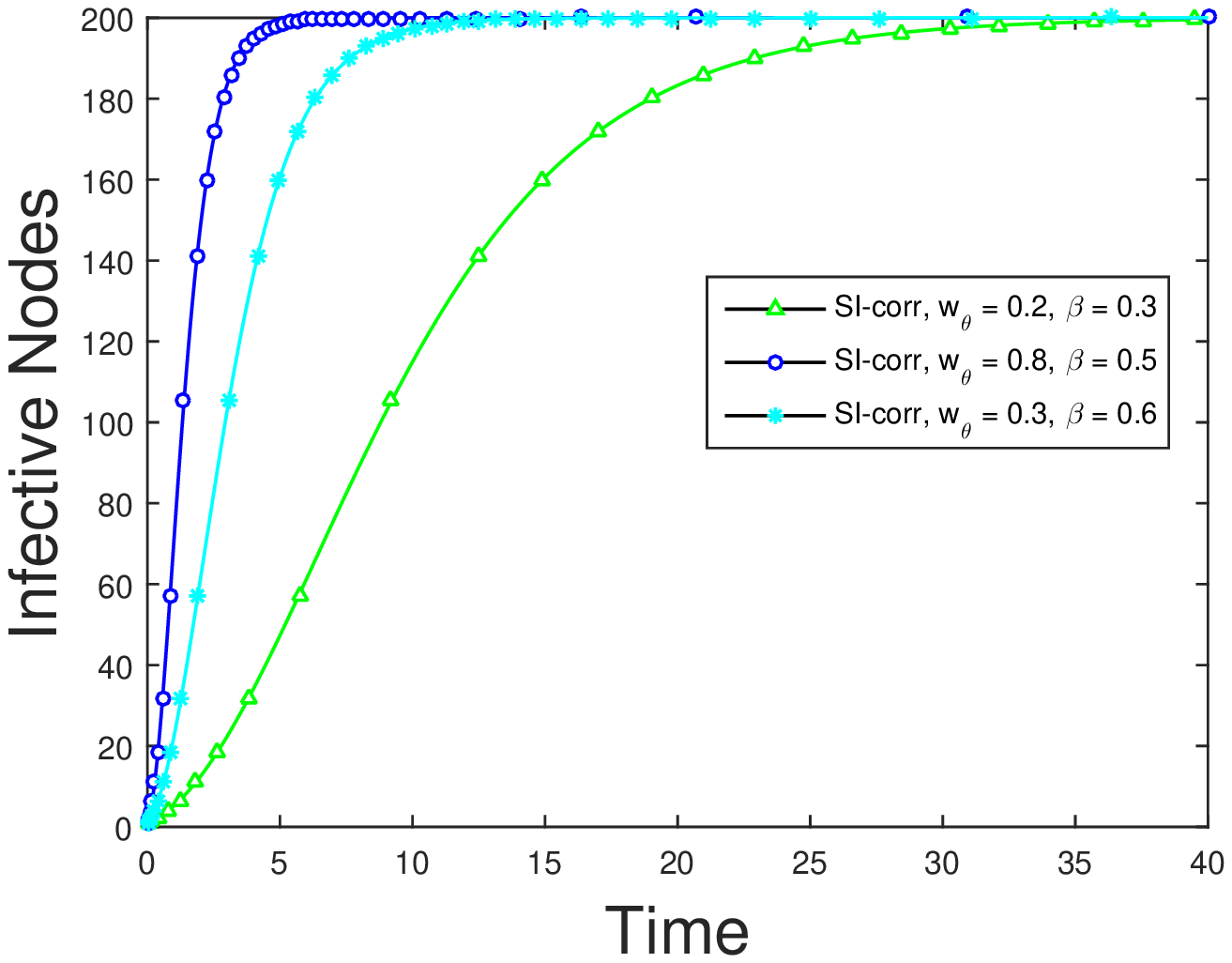} &
\includegraphics[scale=0.50]{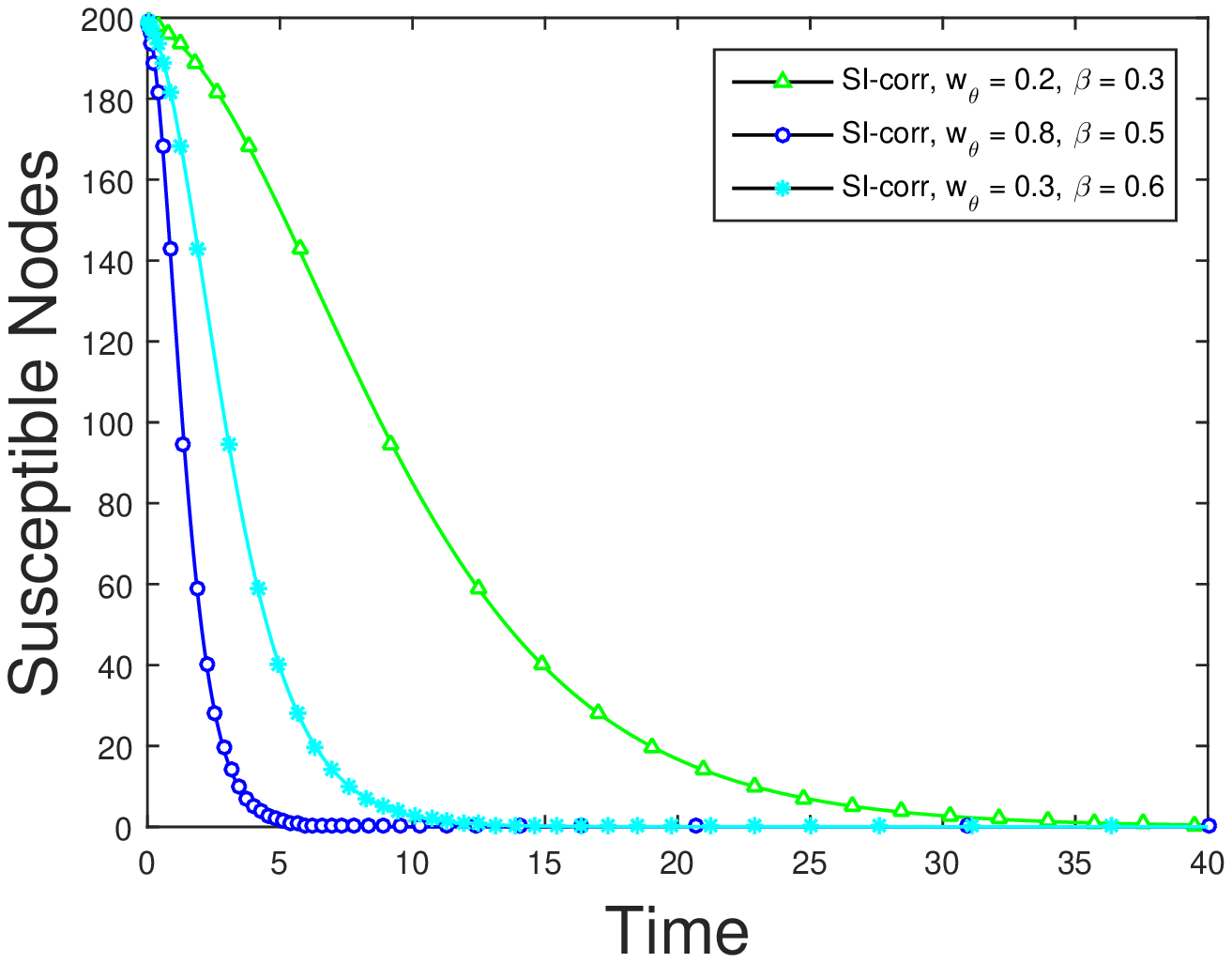}\\ 
\end{tabular}
 \hspace{3in}\parbox{4.8in}{\caption{Time dynamics of infected nodes, I(t) (left figure) and S(t) (right figure) for different values of $w_{\theta}$ (spatial correlation degree) and $\beta$ (infection capacity) for constant values, $N=200$, $r_t=2$, $\sigma=0.5$.}}
\end{center}
\end{figure}

\subsection{Simulation study with Correlated Clustering algorithm}
In this section, we present the virus propagation analysis of our model on spatially correlated algorithm. Simulation setup and implementation details are discussed to capture the time dynamics of the virus spread in WSNs. We have used NS-2 Network Simulator ~\cite{ns2}, which is open-source event-driven simulator for communication protocol design in computer networking field. Total $N=1000$ nodes are deployed uniformly in 2-dimensional field. The fault parameters used for simulation are given in Table~\ref{tabSim}. The CSMA/CA simple MAC scheme is used to transmit the data from one node to another within its transmission range. A false event is created randomly to initiate the virus infection in the network. The evaluating metrics is the time taken by the virus for infecting the fraction of the network with correlation and without correlation. 

In the section~\ref{Lsp}, we discussed that the spatial correlation between nodes is expressed by fraction of overlapped sensing area of nodes of $r_s$-radius disk centered at position of itself. So, $\rho_{(i,j)}$ gives the fraction of overlapped sensing area of nodes of $r_s$-radius disk with $d_{(i,j)}$ separation distance between them. Using $\rho_{(i,j)}$, we define two set of sensor nodes based on their locations and overlapped sensing area. These are highly correlated set (HCS) and weakly correlated set (WCS). 

\begin{table}
\caption{Simulation Parameters}
\label{tabSim}
\centering
\begin{tabular}{|l|p{3.1cm}|l|c|}
\hline
ns2 energy model & $EnergyModel$ & Event sensing range & 400 unit \\ \hline
Initial power     & 100 J& Transmission range  & 200 unit \\ \hline
Tx power & $66$ mW& Event generation interval  & 50 s \\ \hline
Rx power & $39$ mW&Carrier Sensing range  & 550 unit  \\ \hline
Idle power & $35$ mW&Size of RTS/CTS/ACK  & 10 B  \\  \hline
Bandwidth & 20 Kbps& Correlation radius  & 50 m \\  \hline
DIFS & 10ms &  Size of Data  & 100 B  \\ \hline
SIFS & 5ms & Contention Window  & 64 ms  \\ \hline
Retry Limit & 5& Duty Cycle & 5 \% \\ \hline
\end {tabular}
\end{table}

For given node density and number of nodes, a correlation threshold $\xi$ ($0<\xi \leq 1$) can be defined. Using our spatial correlation model, it can be determined that Larger the overlap area between nodes, stronger will be spatial correlation between them~\cite{ref5}. If $\rho_{(i,j)} \geq \xi$, then node $n_i$ and node $n_j$ are strongly correlated. So, the set of nodes is denoted by Highly Correlated Set (HCS). If $\rho_{(i,j)} < \xi$, then node $n_i$ and node $n_j$ are weakly correlated. So, the set of nodes is denoted by Weakly Correlated Set (WCS). Following equalities hold. In~\cite{ref7d}, an efficient clustering approach is proposed that can be used to control the virus spread when infective nodes cross the threshold or limit. For highly correlated set (HCS), $\rho_{(i,j)} \geq \xi$. For weakly correlated set (WCS), $\rho_{(i,j)} < \xi$. When correlation is $\xi$, $d_{(i,j)}=r_{cc}$, i.e.
\begin{equation}
\begin{split}
\label{eq:cor16}
 \xi =	\frac{\cos^{-1}{({r_{cc}}/{2r_s})}}{\pi} -
\frac{r_{cc}}{4\pi{r_s}^2}.\sqrt{(4r_s^2-r_{cc}^2)},
\end{split}
\end{equation} 
For $d_{(i,j)}<r_{cc}$, $\rho_{(i,j)} \geq \xi$. 

Using Eq.~\eqref{eq:cor16}, if any node $n_k$ lying within $r_{cc}$ from node $n_i$, it comes in HCS. Otherwise, set of nodes is WCS, where $r_{cc}$ is denoted as correlation cluster for a given value of $\xi$. To determine the HCS and WCS, Algorithm 1 can be used where set of correlated nodes is denoted by $C = \{C_1,C_2,...\}$ and the formed cluster radius of each set $C$ is denoted by $R_{CC} = \{r_{cc1},r_{cc2},...\}$. This algorithm gives the different values of correlation clusters (i.e., $r_{cc}$) according to the distribution of nodes in the two-dimensional field. The sample results are shown in Table~\ref{tab:2}. This table shows the different values of $\xi$ for given different values of $r_t$. Based on these observations, the actual behavior of virus spreading dynamics in network is investigated. 

\begin{algorithm}
  \algsetup{indent=1.5em}
\caption{Correlation set construction algorithm}
 \begin{algorithmic}
 \STATE $C =
CorrelationSet(\{n_1,n_2,...,n_{N}\},(\rho_{ij})_{N*N},\xi)$
\STATE $R_{CC} =
CorrelationRadiusSet(\{n_1,n_2,...,n_{N}\},(\rho_{ij})_{N*N},\xi)$
\STATE $\textbf{begin}$
 \STATE $\mathcal{N}=\{n_1,n_2,...,n_{N}\},\rho
(n_i,n_j)=\rho_{ij}.$
\FOR {$k=1$ to $N-1$} 
  \STATE Find \begin{equation}CorrelatedPair\{n_i, n_j\} = \argmax_{n_i,n_j \in \mathcal{N}}{\{\rho(n_i,n_j)\}} \nonumber \end{equation} 
 \STATE  $\{$ Find the most correlated pair of set in $\mathcal{N}.\}$
  \STATE  Merge $n_i$ and $n_j$ into new correlation set.
  \STATE  $ C_{i+j} = \{n_i,n_j\}$
	      \FOR {$n_l \in \mathcal{N},l \notin i,l \notin j$} 
 \STATE              Compute $\rho (C_{i+j},n_l)$ and compare it with
$\xi$
\STATE               Add the $n_l$ into set $C_{i+j}$ if more then
$\xi$.
	      \ENDFOR 
 \STATE   Remove $ n_i$ and $n_j$ from $\mathcal{N}$; Define the new
correlation set into $\mathcal{N}$.
 \ENDFOR
 \RETURN $C = \{C_1,C_2,...\}$ and $R_{CC} = \{r_{cc1},r_{cc2},...\}$
\STATE $\textbf{end}$
 \end{algorithmic}
 \end{algorithm}

\begin{table}
\begin{center}
\caption{Results of formed correlated clusters $r_{cc}$ for different values of $r_t$ and $\xi$}
\begin{tabular}{|c|*{5}{c|}}\hline
&\parbox{27.0pt}{$\xi=0.2$}&\parbox{27.0pt}{$\xi=0.4$}&\parbox{27.0pt}{$\xi=0.5$}
&\parbox{27.0pt}{$\xi=0.6$}&\parbox{27.0pt}{$\xi=0.8$}\\\hline
$r_t=9$&\parbox{27.0pt}{5.78}&\parbox{27.0pt}{4.11}&\parbox{27.0pt}{3.51}&\parbox{
27.0pt
} { 2.3}&\parbox{27.0pt}{1.25}\\\hline
$r_t=12$
&\parbox{27.0pt}{8.03}&\parbox{27.0pt}{5.5}&\parbox{27.0pt}{4.7}&\parbox{27.0pt}
{3.05}&\parbox{27.0pt}{1.7}\\\hline
$r_t=15$&\parbox{27.0pt}{10.04}&\parbox{27.0pt}{6.23}&\parbox{27.0pt}{5.87}&\parbox{
27.0pt }
{3.86}&\parbox{27.0pt}{2.16}\\\hline
$r_t=18$
&\parbox{27.0pt}{12.07}&\parbox{27.0pt}{8.23}&\parbox{27.0pt}{7.05}&\parbox{
27.0pt }
{4.67}&\parbox{27.0pt}{2.59}\\\hline
$r_t=21$
&\parbox{27.0pt}{14.0}&\parbox{27.0pt}{9.6}&\parbox{27.0pt}{8.23}&\parbox{27.0pt
}
{5.44}&\parbox{27.0pt}{3.02}\\\hline
$r_t=24$
&\parbox{27.0pt}{16.13}&\parbox{27.0pt}{11.07}&\parbox{27.0pt}{9.39}&\parbox{
27.0pt
}
{6.22}&\parbox{27.0pt}{3.46}\\\hline
\end{tabular}
\label{tab:2}
\end{center}
\end{table}

\end{document}